\documentclass[fleqn]{article} 
\usepackage{graphicx,url}
\usepackage[utf8x]{inputenc}
\usepackage{longtable,color}
\usepackage[top=3cm, bottom=3cm, left=3cm, right=2cm]{geometry}
\usepackage{verbatim}
\usepackage{color}
\usepackage{xspace}
\usepackage{rotating}
\usepackage{times}
\usepackage{latexsym}
\usepackage{amstext}
\usepackage{enumerate}
\usepackage{hyperref}
\usepackage{booktabs} 
\usepackage{amssymb}

\usepackage[utf8x]{inputenc}
\usepackage{color}
\definecolor{codegreen}{rgb}{0,0.4,0}
\definecolor{codeblue}{rgb}{0,0.5,1}
\definecolor{codegray}{rgb}{0.5,0.5,0.5}
\definecolor{codepurple}{rgb}{0.58,0,0.82}
\definecolor{backcolour}{rgb}{0.95,0.95,0.92}

\usepackage[noend, vlined, linesnumbered, ruled, english]{algorithm2e}

\usepackage[scaled=0.9]{DejaVuSansMono}
\usepackage[T1]{fontenc}

\title{Toward a Framework for Integrative, FAIR, and\\Reproducible Management of Data on the\\Dynamic Balance of Microbial Communities}

\author{Luiz Gadelha$^1$\footnote{E-mail address: {\tt luiz.gadelha@uni-jena.de}}, Martin Hohmuth$^1$, Mahnoor Zulfiqar$^1$, David Schöne$^1$, Sheeba Samuel$^1$,\\Maria Sorokina$^2$, Christoph Steinbeck$^1$, Birgitta König-Ries$^1$ 
}
\date{\small
$\mbox{}^1$Friedrich-Schiller University Jena, 07743 Jena, Germany\\
$\mbox{}^2$Research \& Development, Bayer AG, 13353 Berlin, Germany\\
}

\begin{document}

\flushbottom
\maketitle

\begin{abstract}
The increasing volumes of data produced by high-throughput  instruments coupled with advanced computational infrastructures for scientific computing have enabled what is often called a {\em Fourth Paradigm} for scientific research based on the exploration of large datasets.
Current scientific research is often interdisciplinary, making data integration a critical technique for combining data from different scientific domains. Research data management is a critical part of this paradigm, through the proposition and development of methods, techniques, and practices for managing scientific data through their life cycle. Research on microbial communities follows the same pattern of production of large amounts of data obtained, for instance, from sequencing organisms present in environmental samples. Data on microbial communities can come from a multitude of sources and can be stored in different formats. For example, data from metagenomics, metatranscriptomics, metabolomics, and biological imaging are often combined in studies. In this article, we describe the design and current state of implementation of an integrative research data management framework for the Cluster of Excellence Balance of the Microverse aiming to allow for data on microbial communities to be more easily discovered, accessed, combined, and reused. This framework is based on research data repositories and best practices for managing workflows used in the analysis of microbial communities, which includes recording provenance information for tracking data derivation.
\end{abstract}

\noindent {\bf Keywords:} research data management, data integration, computational reproducibility, microbial communities.

\section{Introduction}

Current infrastructures for computational research and the increasing volumes of data produced by high-capacity scientific instruments (e.g. next-generation sequencing and high-end microscopes) have enabled a new way of conducting scientific research, often called data-intensive scientific computing \cite{hey_fourth_2009}. In this context, it is critical to collect and combine data that can come from various sources to pose and explore research questions and hypotheses. Therefore, data integration \cite{miller_open_2018} techniques are essential in enabling this form of research practice.  Also, Mons \cite{mons_data_2018} observed that researchers can spend up to 60\% of their work time on searching, gathering, reformatting, and integrating data for meta-analysis. Data stewardship, which consists of ensuring the quality and fitness-for-use of datasets and associated metadata, should allow for researchers to spend less time doing data wrangling and more time doing actual data analysis. producing machine-readable (or machine-actionable) research data should be a first-order priority. Current research is increasingly done following the {\em open science} approach, which makes {\em research objects} \cite{bechhofer_why_2013}, such as data, scientific software, scientific workflows, and the resulting articles, openly accessible to society. Research objects such as data, methods, protocols, and workflows should follow the FAIR principles \cite{wilkinson_fair_2016}, providing rich and standardized metadata, unique identifiers, and provenance information \cite{herschel_survey_2017}.

Microbial data present similar challenges since they also involve large-scale data collection from high-throughput instruments and are highly heterogeneous in format. 
Kyrpides et al. \cite{kyrpides_microbiome_2016} explored the landscape microbiome data science. They observed that reuse of microbiome data is very limited, often being stored and managed in specialized ways that hinder data integration and meta-analysis. This has negative consequences, such as redundant work. Another point mentioned is the limited availability of reference microbiome data. They recommend the creation of a center for high performance analysis and integration of microbiome data that would explore the comparative analysis of these datasets and ensure their quality to foster better data integration.
Jurburg et al. \cite{jurburg_archives_2020} surveyed the availability of microbial community data by analyzing the existing literature. Text from the articles mentioning 16S rRNA amplicon sequencing was extracted and analyzed with respect to data availability. Even though a small percentage of the works did not make the data public, about 40\% of the data was neither available nor reusable.  In this article, we describe our progress toward a comprehensive integrative research data management framework for the Cluster of Excellence Balance of the Microverse (\url{https://www.microverse-cluster.de}), funded by the German Research Foundation. It is comprised of ten research institutions and about forty research groups focusing on the dynamic balance of microbial communities. 
A framework for the integrative data management of microbial community data was designed and is being implemented encompassing the data repositories BEXIS2 \cite{chamanara_bexis2_2021}, as a general repository for managing microbial community data, and OMERO \cite{allan_omero_2012}, a specialized repository for biological imaging data. A dataset catalogue is also being implemented based on BEXIS2 for aggregating metadata about Microverse datasets that were published on public research data repositories. Additional components include libraries and tools \cite{samuel_collaborative_2022} \cite{mondelli_capturing_2021} \cite{soiland-reyes_packaging_2022} for supporting the computational reproducibility \cite{freire_provenance_2018} \cite{samuel_understanding_2021} of scripts used for analyzing microbial community data. Finally, an instance of the Data Stewardship Wizard \cite{pergl_data_2019} is used for data management planning. 
This framework aims to make research objects, produced across the Microverse cluster, more findable, accessible, interoperable, and reusable, therefore providing good support for the FAIR principles \cite{wilkinson_fair_2016}. 

This article is organized as follows. In section \ref{relwork-sec}, we give an overview of other articles that somehow approach similar problems. In section \ref{impl-sec}, we describe the design and current state of the implementation of the framework. In section \ref{eval-sec}, we provide an evaluation of the current prototype. Finally, in section \ref{concl-sec} we explore the next steps in the development and deployment of the framework and make some remarks.

\section{Related Work}\label{relwork-sec}

Vangay et al. \cite{vangay_microbiome_2021} report on a National Microbiome Data Collaborative (NMDC) workshop that focused on encouraging microbiome data sharing and reducing the barriers for metadata submission. They recognize the work of the Genomics Standards Consortium (GSC) \cite{field_genomic_2011} and the Open Biomedical Ontologies (OBO) Foundry \cite{jackson_obo_2021} in the proposition of sample metadata standards. They surveyed the use of the Minimum Information about any (x) Sequence (MIxS) \cite{yilmaz_minimum_2011} standard in both EMBL and NCBI repositories and observed different patterns of usage of MIxS packages, with ENA focusing on using them more often across studies than across samples, when compared to NCBI. 
Mayer et al. \cite{mayer_implementing_2021} describe six use cases in the German Network for Bioinformatics Infrastructure (de.NBI) and propose them as blueprints for FAIR data management, including practicing metadata collection and general research data management. A self-assessment of the use cases according to the FAIR principles is also provided. 
Eloe-Fadrosh et al. \cite{eloe-fadrosh_national_2022} describe the NMDC Data Portal which allows for discovery and analysis following the FAIR principles. It is built on top of a data schema structured according to the existing microbiome data management practices, linking biosamples, annotations, environmental context, and multi-omic workflows. Data can be explored using, for instance, geographical, environmental, and temporal views.
Cernava et al. \cite{cernava_metadata_2022} propose new efforts for the harmonization  of metadata standards to facilitate large-scale microbiome meta-analysis, including the metadata submission tool recognized globally by public repositories and global minimum required metadata standards. 
 
Our framework for research data management in the Microverse cluster differs from the research mentioned in this section in different ways. First, our cluster does not focus only on microbiome sequencing data, other types of data such as mass spectrometry and biological imaging are also part of the studies. Also, our data management efforts are directed toward early stages of the research life cycle. As we describe in section \ref{impl-sec}, many datasets at later stages of this life cycle are published directly to public repositories, which often delays metadata recording to this data publication stage. By focusing on early stages of research, metadata can be provided during or right after data collection procedures facilitating both early data integration efforts and publication to public repositories.

\section{Design and Current Status of Implementation}\label{impl-sec}

\begin{figure*}
\begin{center}
\includegraphics[width=16cm]{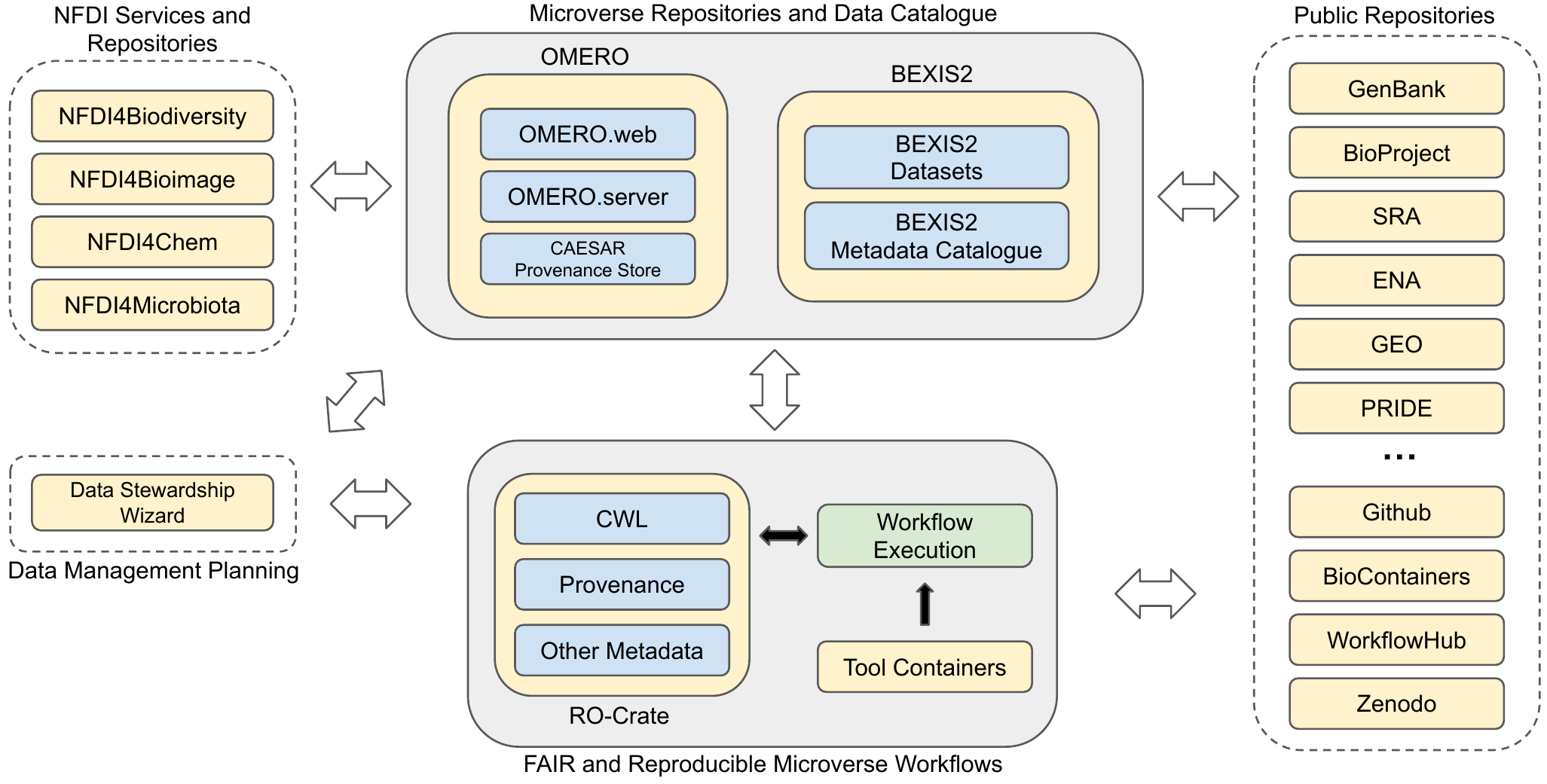}
\caption{Microverse RDM components overview.\label{microverserdm-fig}}
\end{center}
\end{figure*}

A prototype for the integrative data management framework in the cluster was implemented and is continually being improved. It consists of the data repositories BEXIS2 \cite{chamanara_bexis2_2021} and OMERO \cite{allan_omero_2012}, a specialized repository for biological imaging data. BEXIS2 will also work as a dataset catalog by aggregating metadata about datasets that were published on public research data repositories and were produced in the Microverse cluster. Additional components include libraries and tools \cite{samuel_collaborative_2022} \cite{mondelli_capturing_2021} \cite{soiland-reyes_packaging_2022} for supporting the computational reproducibility and scripts used for analyzing microbial community data. Finally, an instance of the Data Stewardship Wizard \cite{pergl_data_2019} is used for data management planning. An overview of the framework is presented in Figure \ref{microverserdm-fig}.

The Microverse cluster comprises a large number of independent groups at different organizations, all with their respective preexisting practices regarding data management. Therefore, a first challenge was to elaborate data management plans (DMPs) in collaboration with stakeholders from the various research groups comprising the cluster. It encompassed activities such as identifying data that needs to be collected, how they will be documented (e.g., using metadata standards), which data quality control techniques will be applied, and how data will be stored, preserved, and disseminated.  Data stewards were appointed in each subgroup of the Microverse cluster and a survey was conducted with them to map current data types, yearly estimates for data size, metadata adoption, compliance to FAIR principles, tools used for data analysis, data retention policies, and repositories used for data publication. The survey showed low adoption of metadata standards in the cluster, which is essential for enabling data integration. Therefore, one of the activities underway is surveying and choosing metadata standards and documenting existing data using best practices and minimum information guidelines. With the support of the Data Stewardship Wizard \cite{pergl_data_2019}, this information was consolidated into DMPs that document and guide research data management in the Microverse cluster. These guidelines include recommendations for research data repositories to which datasets can be published, depending on their type and format. 
Next, we describe two repositories that are part of our framework and aim to support data management for ongoing research in the Microverse cluster, before eventual publication to a domain-specific public repository.


BEXIS2 \cite{chamanara_bexis2_2021} is a general research data repository able to manage different steps in the data life cycle, such as collection, documentation, and preservation. It was developed and is maintained by our research group. Standard metadata profiles, mainly from the biodiversity domain, are supported by default, and metadata schema for other domains can be conveniently added  through its user interface. Additionally, data structures can be defined for tabular data, allowing the automated generation of templates for data collection. Other features of interest include support for data versioning, data linking, authentication and fine-grained authorization, data access through application programming interfaces (e.g. from analysis scripts), and publication of datasets to reference domain repositories. In the Microverse cluster, support for linking datasets in BEXIS2 is leveraged to keep track of the data management workflow. To illustrate this feature, one can add a metadata entry describing a study. Subsequently, one can add metadata for samples and link them to the respective study. If these samples are analyzed and tabular data is derived from them, they can be published in BEXIS2 (along with metadata) and linked to the original samples. This allows for keeping track of the different steps of the research process and recording the relationships among the existing research objects. These relationships may also describe the provenance \cite{oliveira_provenance_2018} of these datasets using ontologies from the W3C PROV \cite{missier_w3c_2013} family of standards, keeping track of data derivation. 


OMERO \cite{allan_omero_2012} is a biological imaging research repository software. An instance of it was installed to store microscopy images collected at the Microverse Imaging Facility, that serves the cluster. Metadata is provided in the OME-XML \cite{goldberg_open_2005} format and additional metadata will be provided using minimum metadata guidelines and the OME.mde extension \cite{kunis_omeromde_2021}. OMERO provides an API that allows for various clients to connect to the repository and explore the datasets available. For instance, Jupyter notebooks can connect to the repository though its Python API. The CAESAR provenance store \cite{samuel_collaborative_2022}, displayed in figure \ref{microverserdm-fig} as a component of OMERO, is a module for capturing provenance information from computational notebooks used for image analysis and was developed for a previous project by our research group. It will be integrated into the Microverse OMERO repository. 


The framework will also benefit from the German National Research Data Infrastrucure (NFDI) \footnote{\url{https://www.nfdi.de/?lang=en}}, which is advancing in providing large-scale research data management services and resources in different scientific domains. NFDI4Chem \cite{steinbeck_nfdi4chem_2020} will provide a virtual environment of federated repositories for managing chemical data, develop minimum information metadata standards, foster the use of electronic laboratory notebooks, and promote open research data management and the FAIR principles \cite{wilkinson_fair_2016}. By fostering the use of ELNs, NFDI4Chem aims to capture data earlier in its life cycle. This is motivated by the fact that currently efficient systems for the curation of user-provided descriptive and contextual data and connection to related device-captured experimental and analytical data are missing. For NFDI4Chem, software for spectroscopic data is also needed since this type of data is currently processed manually or with stand-alone software. NFDI4Chem has developed and deployed Chemotion \cite{tremouilhac_chemotion_2017}, an electronic laboratory notebook tailored for chemical data that we plan to integrate into our framework so that research objects from both infrastructures can be linked to each other. For instance, a notebook in Chemotion could eventually reference a dataset stored either on the BEXIS or OMERO Microverse repositories. The Microverse Data Catalogue could also reference notebooks available on Chemotion from Microverse researchers. Similar integrations are planned with other NFDI initiatives for scientific domains related to the Microverse, such as the 
NFDI4Biodiversity \cite{glockner_nfdi4biodiversity_2019}, which serves the biodiversity, ecology, and environmental data communities, and specially the NFDI4Microbiota \cite{reimer_besser_2022}, which is being implemented to serve the microbiome data community. Finally, the NFDI4BIOIMAGE \cite{schmidt_nfdi4bioimage_2022} was recently recommended for funding, it will serve the biological imaging community.


Hu et al. \cite{hu_challenges_2022} described challenges for managing large-scale microbiome analysis workflows and the approach used at NMDC for managing them, which included the use of workflows management systems, standard workflow definition languages, preservation of computational environments with containers, and a data schemas for storing workflow details. The field of Cheminformatics, for instance, provides different computational algorithms and software to manipulate and analyze the chemical data obtained from analytical high-throughput techniques such as Mass Spectrometry. The reproducibility of these tools is reliant on whether the reimplementation of the source code provides confirmation of the results. To ensure such reproducibility practices, provenance collection from the code can provide validation while re-implementing the tool \cite{clark_path_2019}.

For supporting the computational reproducibility of scripts used for data analysis in the Microverse cluster, we developed tools for collecting semantically enriched provenance information using the REPRODUCE-ME ontology \cite{samuel_end--end_2022} from the execution of scripts written in Python and R \cite{samuel_collaborative_2022} \cite{mondelli_capturing_2021}, which are scripting languages often used in the cluster. This provenance information is essential for interpreting, validating, and reproducing computational analyses. We plan to couple these provenance collection tools to best practices for managing scientific workflows and scripts, improving our support for the FAIR principles for computational workflows \cite{goble_fair_2020} \cite{perezriverol_scalable_2020} \cite{gruening_recommendations_2019}. These include using the Common Workflow Language (CWL) standard \cite{crusoe_methods_2022} for representing the workflows and improving interoperability. Computational activities can also be described using standardized metadata, such as the EDAM ontology \cite{ison_edam_2013}, for describing bioscientific computational activities and data types, and Bioschemas \cite{gray_bioschemas_2017}, for providing general information on research activities in the life sciences. We will use these metadata standards and the RO-Crate  \cite{soiland-reyes_packaging_2022} format for packaging and publishing the scientific workflows and scripts to repositories. In our experience \cite{mondelli_bioworkbench_2018}, scalability is also an important aspect of bioinformatics workflows. Therefore, we are also examining the performance of workflows and applying implicit parallelization \cite{babuji_parsl_2019} for improving scalability.

\section{Results and Prototype Evaluation} \label{eval-sec}

We used the Data Stewardship Wizard \cite{pergl_data_2019} for preparing DMPs. 
The tool allows for importing data stewardship knowledge models developed for different domains. In our case, we used the Life Sciences DSW Knowledge Model (Knowledge Model ID: {\tt dsw:lifesciences:2.3.0}). A questionnaire is derived from the knowledge model and can be applied to collect the information needed for preparing a DMP. Each of the seven Microverse research subgroups indicated a data steward to participate in the process. The process of going through the questionnaire is complete for six of the seven subgroups. In table \ref{summary}, we present a summary of the information collected so far. The DMPs will be periodically reviewed with the data stewards for evaluation and updates.

\begin{table*}[h]
\begin{center}\small 
\caption{Summary of data management plan questionnaire responses.\label{summary}}
\begin{tabular}{  l  p{10.5cm}  }
  {\bf DMP section} & {\bf Responses (number of subgroups)}\\
\hline	
\hline
  Data types & Biological imaging (6), sequencing (5), chemistry (4), metabolomics and proteomics (3), biological assays (2), clinical records (1), environmental (1), sketches (1)\\
\hline
  Sources of reference data & GenBank (5), UniProt (4), SRA (3), MetaboLights (2), SILVA (2), ENA (1), GEO (1), Metabolomics Workbench (1), UNITE (1)\\
\hline
  Target public repositories & GenBank (3), SRA (3), MetaboLights (3), ENA (1), GEO (1), BioImage Archive (1), IDR (1), figshare (1), BioProject (1), Zenodo (1)\\
\hline 
 Metadata standards and ontologies & \href{https://doi.org/10.25504/FAIRsharing.6xq0ee}{GO} (3), \href{https://doi.org/10.25504/FAIRsharing.zk8p4g}{OME-XML} (3), \href{https://doi.org/10.25504/FAIRsharing.9aa0zp}{MIxS} (2), \href{https://doi.org/10.25504/FAIRsharing.26dmba}{mzML} (2), \href{https://doi.org/10.25504/FAIRsharing.fj07xj}{NCBI Taxonomy} (2), \href{https://doi.org/10.25504/FAIRsharing.azqskx}{EnvO} (1), \href{https://doi.org/10.25504/FAIRsharing.xm7tkj}{nmrCV} (1), \href{https://doi.org/10.25504/FAIRsharing.9qv71f}{SBML} (1), \href{https://doi.org/10.25504/FAIRsharing.2rm2b3}{W3C PROV} (1)\\
\hline
 Electronic laboratory notebooks & Benchling (2), RSpace (2), Chemotion (1), eLabFTW (1)\\
 \hline
 Scripting and workflows & R (3), Python (1), QIIME (1), Snakemake (1), Perl (1)\\
 \hline
 Privacy-sensitive data & Yes (1), No (5)\\
\hline

\hline
\end{tabular}\\[2mm]
\end{center} 
\end{table*}

\begin{figure}
\begin{center}
\includegraphics[width=8cm]{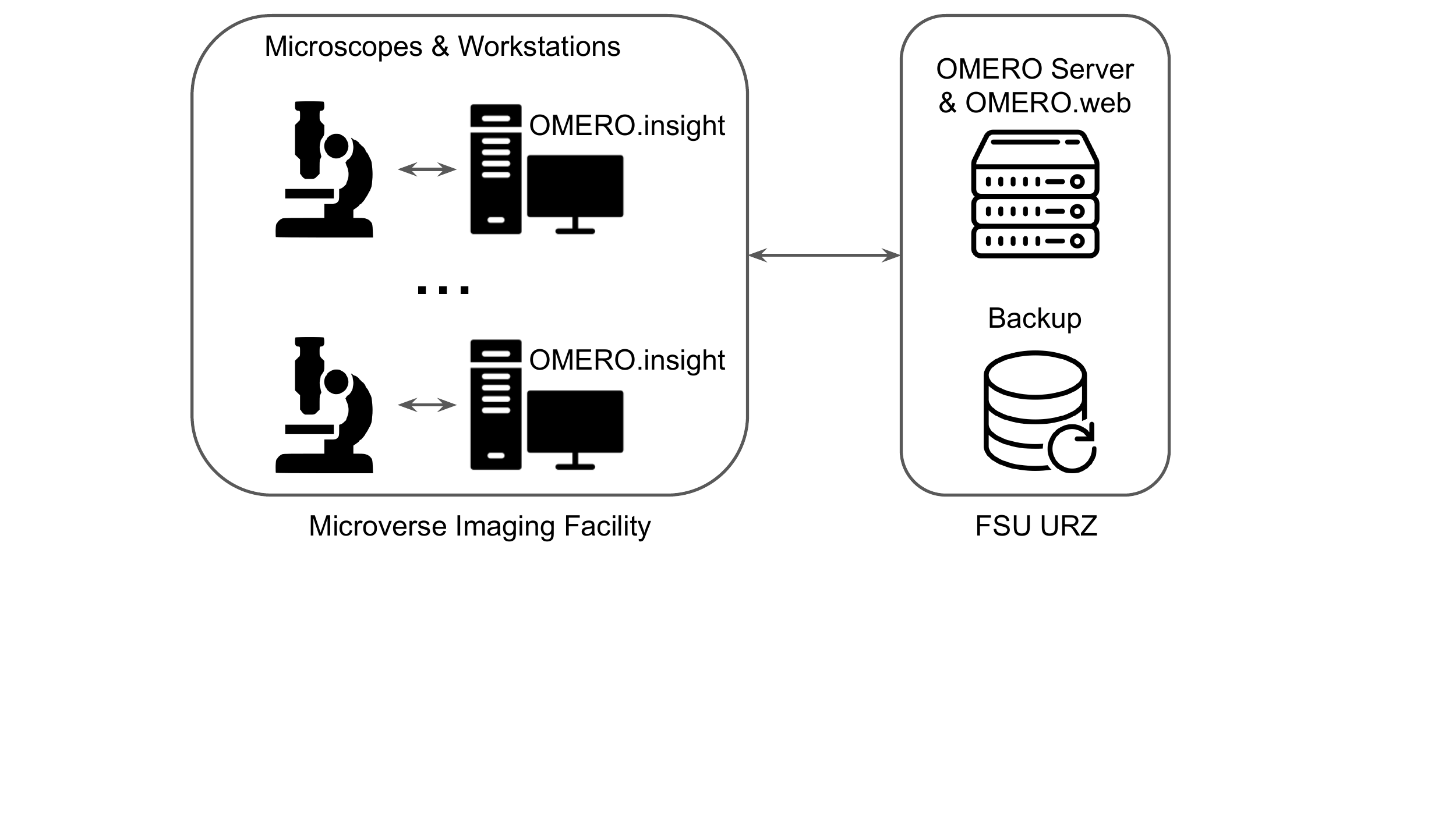}
\caption{Microverse OMERO repository installation overview.\label{omerofig}}
\end{center}
\end{figure}

Two repositories are currently operational and dedicated to managing research data from the Microverse cluster. The Microverse BEXIS2 repository\footnote{\url{http://mv-bexis.bioimbgle.uni-jena.de}} is available and is customized to support the MIxS family of metadata standards in addition to the ones that are pre-installed. Additional metadata standards listed in \ref{summary} are being added. The Microverse OMERO repository\footnote{\url{https://omero.microverse.uni-jena.de/}} was launched recently and as of July 2022 has 16 user accounts and about 36GB of data stored in its early stage of operation. OMERO.insight is a Java-based client that is currently used for uploading datasets to the OMERO repository and can also be used to provide metadata in addition to the microscope-metadata collected automatically. The repository is hosted by the Computing Center of the University of Jena (FSU URZ) and the Microverse Imaging Facility. An overview of the Microverse OMERO repository is provided in figure \ref{omerofig}. 

As a first step toward a dataset catalogue, we surveyed existing Microverse datasets in public research repositories. The procedure consisted of obtaining a list of publications that acknowledge Microverse funding. Each article on the list was examined for data availability sections. For the existing data availability sections, dataset identifiers were collected. Table \ref{survey} shows the distribution of Microverse datasets across different public research repositories as of July 2022. The complete list of datasets and research software can be found on Github\footnote{\url{https://github.com/microverse-rdm/dataset-survey/}} A next step will be to collect metadata from each of these datasets for building the Microverse data catalogue.

\begin{table}[h]
\begin{center}\small 
\caption{Survey of Microverse datasets on public repositories.\label{survey}}
\begin{tabular}{  l  r  }
  {\bf Data type} & {\bf \# of subgroups}\\
\hline	
ArrayExpress &	02\\
BioModels	&		01\\ 
BioProject	&		43\\ 
BMRB	&			06\\ 
Dryad	&			01\\ 
EDMOND	&			02\\ 
ENA	&			10\\ 
Figshare	&		05\\ 
GenBank	(without linked BioProject)&			36\\ 
GEO	&			09\\ 
HKI	&			03\\ 
MassIVE	&			01\\ 	
MetaboLights	&		03\\ 
Metabolomics Workbench	&	01\\ 
OSF	&			02\\ 
PRIDE	&			07\\ 
RefSeq	&			02\\ 
Zenodo	&			03\\ 
\hline
{\bf Total} & 137
\end{tabular}

\end{center} 
\end{table}

We are currently working on applying best practices for supporting reproducibility and the FAIR principles \cite{goble_fair_2020} in the Metabolome Annotation Workflow\footnote{\url{https://github.com/zmahnoor14/MAW}} (MAW), which is a data analysis pipeline for Mass Spectrometry data. So far, we have worked on preserving the computational environments and dependencies of tools that compose the workflow using the Conda package manager and Docker containers. The workflow is being encoded using CWL \cite{crusoe_methods_2022}. 

Next steps include obtaining unique identifiers for the research objects that comprise analyses performed with MAW, providing rich contextual metadata, including provenance \cite{mondelli_capturing_2021}, using appropriate ontologies and metadata standards. Finally the workflow and contextual metadata will be package and published using the RO-Crate \cite{soiland-reyes_packaging_2022} format. With the practices already in place described in this section, we advanced toward providing rich and standardized metadata along with unique identifiers for the research objects in the Microverse cluster, which are accessible through our repositories. Therefore, providing better support for the FAIR principles.

\section{Future Work and Concluding Remarks} \label{concl-sec}

In this article, we described the design and the current state of implementation of a framework for integrative research data management for the Microverse cluster. Our survey of data management workflows and practices in the various research groups that are part of the cluster provided a basis for the DMPs but and supported also the identification and proposition of design patterns for microbial data management. As next steps, we plan to advance with the interconnection of the components described and to propose data management workflows to capture information from study conception and definition to the resulting datasets. 
Many research groups in the Microverse cluster already publish data in reference domain repositories, such as the Sequence Read Archive (SRA) and MetaboLights, for high-throughput sequencing and metabolomics data, respectively. For data integration, it is important to keep track of these datasets as well so they can be incorporated into new studies. Therefore, a data catalog for the cluster is being by harvesting metadata from both the internal and public repositories.

We plan to further integrate the provenance collection tools and libraries described, based on the REPRODUCE-ME ontology, into a computational reproducibility framework previously proposed \cite{mondelli_exploring_2019}. This will allow not only the collection of provenance information from script executions but also the preservation of reproducibility-critical information, such as the description and documentation of the computational environments used for execution. 
With the framework described here, we believe that better support will be provided for following the FAIR principles in microbial data management, not only for data but also for other types of research objects, such as scripts and workflows.
Finally, we plan to explore other data integration techniques in addition to the metadata catalogue. Knowledge graphs  \cite{joachimiak_kg-microbe_2021}, for instance, can be built from information extracted from the academic literature produced by the Microverse cluster and its metadata catalogue. They can enable more sophisticated querying, search, and reasoning about microbial community data.

\section*{Acknowledgments}   
Funded by the Deutsche Forschungsgemeinschaft (DFG, German Research Foundation) under Germany's Excellence Strategy -- EXC 2051 -- Project-ID 390713860 and Deutsche Forschungsgemeinschaft (DFG, German Research Foundation), Project-ID 239748522, SFB 1127 ChemBioSys.


\begin{thebibliography}{10}
\providecommand{\url}[1]{#1}
\csname url@samestyle\endcsname
\providecommand{\newblock}{\relax}
\providecommand{\bibinfo}[2]{#2}
\providecommand{\BIBentrySTDinterwordspacing}{\spaceskip=0pt\relax}
\providecommand{\BIBentryALTinterwordstretchfactor}{4}
\providecommand{\BIBentryALTinterwordspacing}{\spaceskip=\fontdimen2\font plus
\BIBentryALTinterwordstretchfactor\fontdimen3\font minus
  \fontdimen4\font\relax}
\providecommand{\BIBforeignlanguage}[2]{{%
\expandafter\ifx\csname l@#1\endcsname\relax
\typeout{** WARNING: IEEEtran.bst: No hyphenation pattern has been}%
\typeout{** loaded for the language `#1'. Using the pattern for}%
\typeout{** the default language instead.}%
\else
\language=\csname l@#1\endcsname
\fi
#2}}
\providecommand{\BIBdecl}{\relax}
\BIBdecl

\bibitem{hey_fourth_2009}
T.~Hey, S.~Tansley, and K.~Tolle, \emph{The {Fourth} {Paradigm}:
  {Data}-{Intensive} {Scientific} {Discovery}}.\hskip 1em plus 0.5em minus
  0.4em\relax Microsoft Research, 2009.

\bibitem{miller_open_2018}
\BIBentryALTinterwordspacing
R.~J. Miller, ``Open {Data} {Integration},'' \emph{Proceedings of the VLDB
  Endowment}, vol.~11, no.~12, pp. 2130--2139, 2018. [Online]. Available:
  \url{https://doi.org/10.14778/3229863.3240491}
\BIBentrySTDinterwordspacing

\bibitem{mons_data_2018}
B.~Mons, \emph{Data {Stewardship} for {Open} {Science}: {Implementing} {FAIR}
  {Principles}}.\hskip 1em plus 0.5em minus 0.4em\relax CRC Press, 2018.

\bibitem{bechhofer_why_2013}
\BIBentryALTinterwordspacing
S.~Bechhofer, I.~Buchan, D.~De~Roure, P.~Missier, J.~Ainsworth, J.~Bhagat,
  P.~Couch, D.~Cruickshank, M.~Delderfield, I.~Dunlop, M.~Gamble,
  D.~Michaelides, S.~Owen, D.~Newman, S.~Sufi, and C.~Goble, ``Why linked data
  is not enough for scientists,'' \emph{Future Generation Computer Systems},
  vol.~29, no.~2, pp. 599--611, Feb. 2013. [Online]. Available:
  \url{https://doi.org/10.1016/j.future.2011.08.004}
\BIBentrySTDinterwordspacing

\bibitem{wilkinson_fair_2016}
\BIBentryALTinterwordspacing
M.~D. Wilkinson, M.~Dumontier, I.~J. Aalbersberg, G.~Appleton, M.~Axton,
  A.~Baak, N.~Blomberg, J.-W. Boiten, L.~B. da~Silva~Santos, P.~E. Bourne,
  J.~Bouwman, A.~J. Brookes, T.~Clark, M.~Crosas, I.~Dillo, O.~Dumon,
  S.~Edmunds, C.~T. Evelo, R.~Finkers, A.~Gonzalez-Beltran, A.~J. Gray,
  P.~Groth, C.~Goble, J.~S. Grethe, J.~Heringa, P.~A. 't~Hoen, R.~Hooft,
  T.~Kuhn, R.~Kok, J.~Kok, S.~J. Lusher, M.~E. Martone, A.~Mons, A.~L. Packer,
  B.~Persson, P.~Rocca-Serra, M.~Roos, R.~van Schaik, S.-A. Sansone,
  E.~Schultes, T.~Sengstag, T.~Slater, G.~Strawn, M.~A. Swertz, M.~Thompson,
  J.~van~der Lei, E.~van Mulligen, J.~Velterop, A.~Waagmeester, P.~Wittenburg,
  K.~Wolstencroft, J.~Zhao, and B.~Mons, ``The {FAIR} {Guiding} {Principles}
  for scientific data management and stewardship,'' \emph{Scientific Data},
  vol.~3, p. 160018, Mar. 2016. [Online]. Available:
  \url{http://doi.org/10.1038/sdata.2016.18}
\BIBentrySTDinterwordspacing

\bibitem{herschel_survey_2017}
\BIBentryALTinterwordspacing
M.~Herschel, R.~Diestelkämper, and H.~Ben~Lahmar, ``A survey on provenance:
  {What} for? {What} form? {What} from?'' \emph{The VLDB Journal}, vol.~26,
  no.~6, pp. 881--906, Dec. 2017, publisher: Springer Berlin Heidelberg.
  [Online]. Available: \url{https://doi.org/10.1007/s00778-017-0486-1}
\BIBentrySTDinterwordspacing

\bibitem{kyrpides_microbiome_2016}
\BIBentryALTinterwordspacing
N.~C. Kyrpides, E.~A. Eloe-Fadrosh, and N.~N. Ivanova, ``Microbiome {Data}
  {Science}: {Understanding} {Our} {Microbial} {Planet},'' \emph{Trends in
  Microbiology}, vol.~24, no.~6, pp. 425--427, Jun. 2016. [Online]. Available:
  \url{https://doi.org/10.1016/j.tim.2016.02.011}
\BIBentrySTDinterwordspacing

\bibitem{jurburg_archives_2020}
\BIBentryALTinterwordspacing
S.~D. Jurburg, M.~Konzack, N.~Eisenhauer, and A.~Heintz-Buschart, ``The
  archives are half-empty: an assessment of the availability of microbial
  community sequencing data,'' \emph{Communications Biology}, vol.~3, no.~1, p.
  474, Dec. 2020. [Online]. Available:
  \url{https://doi.org/10.1038/s42003-020-01204-9}
\BIBentrySTDinterwordspacing

\bibitem{chamanara_bexis2_2021}
\BIBentryALTinterwordspacing
J.~Chamanara, J.~Gaikwad, R.~Gerlach, A.~Algergawy, A.~Ostrowski, and
  B.~König-Ries, ``{BEXIS2}: {A} {FAIR}-aligned data management system for
  biodiversity, ecology and environmental data,'' \emph{Biodiversity Data
  Journal}, vol.~9, Nov. 2021. [Online]. Available:
  \url{https://doi.org/10.3897/BDJ.9.e72901}
\BIBentrySTDinterwordspacing

\bibitem{allan_omero_2012}
\BIBentryALTinterwordspacing
C.~Allan, J.-M. Burel, J.~Moore, C.~Blackburn, M.~Linkert, S.~Loynton,
  D.~MacDonald, W.~J. Moore, C.~Neves, A.~Patterson, M.~Porter, A.~Tarkowska,
  B.~Loranger, J.~Avondo, I.~Lagerstedt, L.~Lianas, S.~Leo, K.~Hands, R.~T.
  Hay, A.~Patwardhan, C.~Best, G.~J. Kleywegt, G.~Zanetti, and J.~R. Swedlow,
  ``{OMERO}: flexible, model-driven data management for experimental biology,''
  \emph{Nature Methods}, vol.~9, no.~3, pp. 245--253, Mar. 2012. [Online].
  Available: \url{http://www.nature.com/articles/nmeth.1896}
\BIBentrySTDinterwordspacing

\bibitem{samuel_collaborative_2022}
\BIBentryALTinterwordspacing
S.~Samuel and B.~König-Ries, ``A collaborative semantic-based provenance
  management platform for reproducibility,'' \emph{PeerJ Computer Science},
  vol.~8, p. e921, Mar. 2022, publisher: PeerJ Inc. [Online]. Available:
  \url{https://peerj.com/articles/cs-921}
\BIBentrySTDinterwordspacing

\bibitem{mondelli_capturing_2021}
\BIBentryALTinterwordspacing
M.~L. Mondelli, S.~Samuel, B.~Konig-Ries, and L.~M.~R. Gadelha, ``Capturing and
  {Semantically} {Describing} {Provenance} to {Tell} the {Story} of {R}
  {Scripts},'' in \emph{2021 {IEEE} 17th {International} {Conference} on
  {eScience} ({eScience})}.\hskip 1em plus 0.5em minus 0.4em\relax IEEE, Sep.
  2021, pp. 283--288. [Online]. Available:
  \url{https://ieeexplore.ieee.org/document/9582412/}
\BIBentrySTDinterwordspacing

\bibitem{soiland-reyes_packaging_2022}
\BIBentryALTinterwordspacing
S.~Soiland-Reyes, P.~Sefton, M.~Crosas, L.~J. Castro, F.~Coppens, J.~M.
  Fernández, D.~Garijo, B.~Grüning, M.~La~Rosa, S.~Leo, E.~Ó~Carragáin,
  M.~Portier, A.~Trisovic, {RO-Crate Community}, P.~Groth, and C.~Goble,
  ``Packaging research artefacts with {RO}-{Crate},'' \emph{Data Science}, pp.
  1--42, Jan. 2022. [Online]. Available:
  \url{https://doi.org/10.3233/DS-210053}
\BIBentrySTDinterwordspacing

\bibitem{freire_provenance_2018}
\BIBentryALTinterwordspacing
J.~Freire and F.~Chirigati, ``Provenance and the {Different} {Flavors} of
  {Computational} {Reproducibility},'' \emph{Bulletin of the Technical
  Committee on Data Engineering}, vol.~41, no.~1, pp. 15--26, 2018. [Online].
  Available: \url{http://sites.computer.org/debull/A18mar/p15.pdf}
\BIBentrySTDinterwordspacing

\bibitem{samuel_understanding_2021}
\BIBentryALTinterwordspacing
S.~Samuel and B.~König-Ries, ``Understanding experiments and research
  practices for reproducibility: an exploratory study,'' \emph{PeerJ}, vol.~9,
  p. e11140, Apr. 2021. [Online]. Available:
  \url{https://peerj.com/articles/11140}
\BIBentrySTDinterwordspacing

\bibitem{pergl_data_2019}
\BIBentryALTinterwordspacing
R.~Pergl, R.~Hooft, M.~Suchánek, V.~Knaisl, and J.~Slifka, ``“{Data}
  {Stewardship} {Wizard}”: {A} {Tool} {Bringing} {Together} {Researchers},
  {Data} {Stewards}, and {Data} {Experts} around {Data} {Management}
  {Planning},'' \emph{Data Science Journal}, vol.~18, Dec. 2019. [Online].
  Available: \url{http://datascience.codata.org/articles/10.5334/dsj-2019-059/}
\BIBentrySTDinterwordspacing

\bibitem{vangay_microbiome_2021}
\BIBentryALTinterwordspacing
P.~Vangay, J.~Burgin, A.~Johnston, K.~L. Beck, D.~C. Berrios, K.~Blumberg,
  S.~Canon, P.~Chain, J.-M. Chandonia, D.~Christianson, S.~V. Costes,
  J.~Damerow, W.~D. Duncan, J.~P. Dundore-Arias, K.~Fagnan, J.~M. Galazka,
  S.~M. Gibbons, D.~Hays, J.~Hervey, B.~Hu, B.~L. Hurwitz, P.~Jaiswal, M.~P.
  Joachimiak, L.~Kinkel, J.~Ladau, S.~L. Martin, L.~A. McCue, K.~Miller,
  N.~Mouncey, C.~Mungall, E.~Pafilis, T.~B.~K. Reddy, L.~Richardson, S.~Roux,
  L.~M. Schriml, J.~P. Shaffer, J.~C. Sundaramurthi, L.~R. Thompson, R.~E.
  Timme, J.~Zheng, E.~M. Wood-Charlson, and E.~A. Eloe-Fadrosh, ``Microbiome
  {Metadata} {Standards}: {Report} of the {National} {Microbiome} {Data}
  {Collaborative}'s {Workshop} and {Follow}-{On} {Activities},''
  \emph{mSystems}, vol.~6, no.~1, Feb. 2021. [Online]. Available:
  \url{https://journals.asm.org/doi/10.1128/mSystems.01194-20}
\BIBentrySTDinterwordspacing

\bibitem{field_genomic_2011}
\BIBentryALTinterwordspacing
D.~Field, L.~Amaral-Zettler, G.~Cochrane, J.~R. Cole, P.~Dawyndt, G.~M.
  Garrity, J.~Gilbert, F.~O. Glöckner, L.~Hirschman, I.~Karsch-Mizrachi, H.~P.
  Klenk, R.~Knight, R.~Kottmann, N.~Kyrpides, F.~Meyer, I.~S. Gil, S.~A.
  Sansone, L.~M. Schriml, P.~Sterk, T.~Tatusova, D.~W. Ussery, O.~White, and
  J.~Wooley, ``The {Genomic} {Standards} {Consortium},'' \emph{PLOS Biology},
  vol.~9, no.~6, p. e1001088, Jun. 2011, publisher: Public Library of Science.
  [Online]. Available:
  \url{https://journals.plos.org/plosbiology/article?id=10.1371/journal.pbio.1001088}
\BIBentrySTDinterwordspacing

\bibitem{jackson_obo_2021}
\BIBentryALTinterwordspacing
R.~Jackson, N.~Matentzoglu, J.~A. Overton, R.~Vita, J.~P. Balhoff, P.~L.
  Buttigieg, S.~Carbon, M.~Courtot, A.~D. Diehl, D.~M. Dooley, W.~D. Duncan,
  N.~L. Harris, M.~A. Haendel, S.~E. Lewis, D.~A. Natale, D.~Osumi-Sutherland,
  A.~Ruttenberg, L.~M. Schriml, B.~Smith, C.~J. Stoeckert~Jr., N.~A.
  Vasilevsky, R.~L. Walls, J.~Zheng, C.~J. Mungall, and B.~Peters, ``{OBO}
  {Foundry} in 2021: operationalizing open data principles to evaluate
  ontologies,'' \emph{Database}, vol. 2021, Oct. 2021. [Online]. Available:
  \url{https://doi.org/10.1093/database/baab069}
\BIBentrySTDinterwordspacing

\bibitem{yilmaz_minimum_2011}
\BIBentryALTinterwordspacing
P.~Yilmaz, R.~Kottmann, D.~Field, R.~Knight, J.~R. Cole, L.~Amaral-Zettler,
  J.~A. Gilbert, I.~Karsch-Mizrachi, A.~Johnston, G.~Cochrane, R.~Vaughan,
  C.~Hunter, J.~Park, N.~Morrison, P.~Rocca-Serra, P.~Sterk, M.~Arumugam,
  M.~Bailey, L.~Baumgartner, B.~W. Birren, M.~J. Blaser, V.~Bonazzi, T.~Booth,
  P.~Bork, F.~D. Bushman, P.~L. Buttigieg, P.~S.~G. Chain, E.~Charlson, E.~K.
  Costello, H.~Huot-Creasy, P.~Dawyndt, T.~DeSantis, N.~Fierer, J.~A. Fuhrman,
  R.~E. Gallery, D.~Gevers, R.~A. Gibbs, I.~San~Gil, A.~Gonzalez, J.~I. Gordon,
  R.~Guralnick, W.~Hankeln, S.~Highlander, P.~Hugenholtz, J.~Jansson, A.~L.
  Kau, S.~T. Kelley, J.~Kennedy, D.~Knights, O.~Koren, J.~Kuczynski,
  N.~Kyrpides, R.~Larsen, C.~L. Lauber, T.~Legg, R.~E. Ley, C.~A. Lozupone,
  W.~Ludwig, D.~Lyons, E.~Maguire, B.~A. Methé, F.~Meyer, B.~Muegge,
  S.~Nakielny, K.~E. Nelson, D.~Nemergut, J.~D. Neufeld, L.~K. Newbold, A.~E.
  Oliver, N.~R. Pace, G.~Palanisamy, J.~Peplies, J.~Petrosino, L.~Proctor,
  E.~Pruesse, C.~Quast, J.~Raes, S.~Ratnasingham, J.~Ravel, D.~A. Relman,
  S.~Assunta-Sansone, P.~D. Schloss, L.~Schriml, R.~Sinha, M.~I. Smith,
  E.~Sodergren, A.~Spo, J.~Stombaugh, J.~M. Tiedje, D.~V. Ward, G.~M.
  Weinstock, D.~Wendel, O.~White, A.~Whiteley, A.~Wilke, J.~R. Wortman,
  T.~Yatsunenko, and F.~O. Glöckner, ``Minimum information about a marker gene
  sequence ({MIMARKS}) and minimum information about any (x) sequence ({MIxS})
  specifications.'' \emph{Nature biotechnology}, vol.~29, no.~5, pp. 415--20,
  May 2011. [Online]. Available: \url{http://dx.doi.org/10.1038/nbt.1823}
\BIBentrySTDinterwordspacing

\bibitem{mayer_implementing_2021}
\BIBentryALTinterwordspacing
G.~Mayer, W.~Müller, K.~Schork, J.~Uszkoreit, A.~Weidemann, U.~Wittig, M.~Rey,
  C.~Quast, J.~Felden, F.~O. Glöckner, M.~Lange, D.~Arend, S.~Beier,
  A.~Junker, U.~Scholz, D.~Schüler, H.~A. Kestler, D.~Wibberg, A.~Pühler,
  S.~Twardziok, J.~Eils, R.~Eils, S.~Hoffmann, M.~Eisenacher, and M.~Turewicz,
  ``Implementing {FAIR} data management within the {German} {Network} for
  {Bioinformatics} {Infrastructure} (de.{NBI}) exemplified by selected use
  cases,'' \emph{Briefings in Bioinformatics}, Feb. 2021. [Online]. Available:
  \url{https://doi.org/10.1093/bib/bbab010}
\BIBentrySTDinterwordspacing

\bibitem{eloe-fadrosh_national_2022}
\BIBentryALTinterwordspacing
E.~A. Eloe-Fadrosh, F.~A. Anubhav, M.~Babinski, J.~Baumes, M.~Borkum,
  L.~Bramer, S.~Canon, D.~S. Christianson, Y.~E. Corilo, K.~W. Davenport,
  B.~Davis, M.~Drake, W.~D. Duncan, M.~C. Flynn, D.~Hays, B.~Hu, M.~Huntemann,
  J.~Kelliher, S.~Lebedeva, P.-E. Li, M.~Lipton, C.-C. Lo, S.~Martin,
  D.~Millard, K.~Miller, M.~A. Miller, P.~Piehowski, E.~P. Jackson, S.~Purvine,
  T.~B.~K. Reddy, R.~Richardson, M.~Rudolph, S.~Sarrafan, M.~Shakya, M.~Smith,
  K.~Stratton, J.~C. Sundaramurthi, P.~Vangay, D.~Winston, E.~M. Wood-Charlson,
  Y.~Xu, P.~S.~G. Chain, L.~A. McCue, D.~Mans, C.~J. Mungall, N.~J. Mouncey,
  and K.~Fagnan, ``The {National} {Microbiome} {Data} {Collaborative} {Data}
  {Portal}: an integrated multi-omics microbiome data resource,'' \emph{Nucleic
  Acids Research}, vol.~50, no.~D1, pp. D828--D836, 2022. [Online]. Available:
  \url{http://doi.org/10.1093/nar/gkab990}
\BIBentrySTDinterwordspacing

\bibitem{cernava_metadata_2022}
\BIBentryALTinterwordspacing
T.~Cernava, D.~Rybakova, F.~Buscot, T.~Clavel, A.~C. McHardy, F.~Meyer,
  F.~Meyer, J.~Overmann, B.~Stecher, A.~Sessitsch, M.~Schloter, G.~Berg, {The
  MicrobiomeSupport Team}, P.~Arruda, T.~Bartzanas, T.~Kostic, P.~I. Brennan,
  B.~B. Biazotti, M.-C. Champomier-Verges, T.~Charles, M.~Coakley, P.~Cotter,
  D.~Cowan, K.~D’Hondt, I.~Ferrocino, K.~Foterek, G.~Herrero-Corral,
  C.~Huitema, J.~Jansson, S.-J. Liu, P.~Malloy, E.~Maguin, L.~Markiewicz,
  R.~Mcclure, A.~Moser, J.~Roovers, M.~Ryan, I.~Sarand, B.~Schelkle,
  A.~Meisner, U.~Schurr, J.~Selvin, E.~Tsakalidou, M.~Wagner, S.~Wakelin,
  W.~Wiczkowski, H.~Winkler, J.~Xiao, C.~J. Bunthof, R.~S.~C. de~Souza,
  Y.~Sanz, L.~Lange, and H.~Smidt, ``\BIBforeignlanguage{en}{Metadata
  harmonization–{Standards} are the key for a better usage of omics data for
  integrative microbiome analysis},''
  \emph{\BIBforeignlanguage{en}{Environmental Microbiome}}, vol.~17, no.~1,
  p.~33, Dec. 2022. [Online]. Available:
  \url{https://doi.org/10.1186/s40793-022-00425-1}
\BIBentrySTDinterwordspacing

\bibitem{oliveira_provenance_2018}
\BIBentryALTinterwordspacing
W.~Oliveira, D.~D. Oliveira, and V.~Braganholo, ``Provenance {Analytics} for
  {Workflow}-{Based} {Computational} {Experiments},'' \emph{ACM Computing
  Surveys}, vol.~51, no.~3, pp. 1--25, May 2018. [Online]. Available:
  \url{http://dl.acm.org/citation.cfm?doid=3212709.3184900}
\BIBentrySTDinterwordspacing

\bibitem{missier_w3c_2013}
\BIBentryALTinterwordspacing
P.~Missier, K.~Belhajjame, and J.~Cheney, ``The {W3C} {PROV} family of
  specifications for modelling provenance metadata,'' in \emph{Proceedings of
  the 16th {International} {Conference} on {Extending} {Database} {Technology}
  - {EDBT} '13}.\hskip 1em plus 0.5em minus 0.4em\relax New York, New York,
  USA: ACM Press, Mar. 2013, p. 773. [Online]. Available:
  \url{http://dl.acm.org/citation.cfm?id=2452376.2452478}
\BIBentrySTDinterwordspacing

\bibitem{goldberg_open_2005}
\BIBentryALTinterwordspacing
I.~G. Goldberg, C.~Allan, J.-M. Burel, D.~Creager, A.~Falconi, H.~Hochheiser,
  J.~Johnston, J.~Mellen, P.~K. Sorger, and J.~R. Swedlow, ``The {Open}
  {Microscopy} {Environment} ({OME}) {Data} {Model} and {XML} file: open tools
  for informatics and quantitative analysis in biological imaging.''
  \emph{Genome biology}, vol.~6, no.~5, p. R47, 2005. [Online]. Available:
  \url{https://doi.org/10.1186/gb-2005-6-5-r47}
\BIBentrySTDinterwordspacing

\bibitem{kunis_omeromde_2021}
\BIBentryALTinterwordspacing
S.~Kunis, S.~Hänsch, C.~Schmidt, F.~Wong, and S.~Weidtkamp-Peters,
  ``{OMERO}.mde in a use case for microscopy metadata harmonization:
  {Facilitating} {FAIR} principles in practical application with metadata
  annotation tools,'' Mar. 2021, \_eprint: 2103.02942. [Online]. Available:
  \url{http://arxiv.org/abs/2103.02942}
\BIBentrySTDinterwordspacing

\bibitem{steinbeck_nfdi4chem_2020}
\BIBentryALTinterwordspacing
C.~Steinbeck, O.~Koepler, F.~Bach, S.~Herres-Pawlis, N.~Jung, J.~Liermann,
  S.~Neumann, M.~Razum, C.~Baldauf, F.~Biedermann, T.~Bocklitz, F.~Boehm,
  F.~Broda, P.~Czodrowski, T.~Engel, M.~Hicks, S.~Kast, C.~Kettner, W.~Koch,
  G.~Lanza, A.~Link, R.~Mata, W.~Nagel, A.~Porzel, N.~Schlörer, T.~Schulze,
  H.-G. Weinig, W.~Wenzel, L.~Wessjohann, and S.~Wulle, ``{NFDI4Chem} -
  {Towards} a {National} {Research} {Data} {Infrastructure} for {Chemistry} in
  {Germany},'' \emph{Research Ideas and Outcomes}, vol.~6, Jun. 2020. [Online].
  Available: \url{https://riojournal.com/article/55852/}
\BIBentrySTDinterwordspacing

\bibitem{tremouilhac_chemotion_2017}
\BIBentryALTinterwordspacing
P.~Tremouilhac, A.~Nguyen, Y.-C. Huang, S.~Kotov, D.~S. Lütjohann, F.~Hübsch,
  N.~Jung, and S.~Bräse, ``Chemotion {ELN}: an {Open} {Source} electronic lab
  notebook for chemists in academia,'' \emph{Journal of Cheminformatics},
  vol.~9, no.~1, p.~54, Dec. 2017. [Online]. Available:
  \url{https://jcheminf.biomedcentral.com/articles/10.1186/s13321-017-0240-0}
\BIBentrySTDinterwordspacing

\bibitem{glockner_nfdi4biodiversity_2019}
\BIBentryALTinterwordspacing
F.~O. Glöckner and M.~Diepenbroek, ``{NFDI4BioDiversity}: {Biodiversity},
  ecology and environmental data,'' \emph{Biodiversity Information Science and
  Standards}, vol.~3, Jul. 2019. [Online]. Available:
  \url{https://biss.pensoft.net/article/37282/}
\BIBentrySTDinterwordspacing

\bibitem{reimer_besser_2022}
\BIBentryALTinterwordspacing
L.~C. Reimer, K.~U. Förstner, and J.~Overmann,
  ``\BIBforeignlanguage{de}{Besser forschen durch offene und {FAIRe}
  {Daten}},'' \emph{\BIBforeignlanguage{de}{BIOspektrum}}, vol.~28, no.~2, pp.
  223--223, Mar. 2022. [Online]. Available:
  \url{https://link.springer.com/10.1007/s12268-022-1725-6}
\BIBentrySTDinterwordspacing

\bibitem{schmidt_nfdi4bioimage_2022}
\BIBentryALTinterwordspacing
C.~Schmidt and E.~Ferrando-May, ``{NFDI4BIOIMAGE} – {An} {Initiative} for a
  {National} {Research} {Data} {Infrastructure} for {Microscopy} {Data},'' in
  \emph{E-{Science}-{Tage} 2021. 339 {Share} {Your} {Research} {Data}}, 2022,
  publisher: heiBOOKS. [Online]. Available:
  \url{https://books.ub.uni-heidelberg.de/index.php/heibooks/catalog/book/979/c13747}
\BIBentrySTDinterwordspacing

\bibitem{hu_challenges_2022}
\BIBentryALTinterwordspacing
B.~Hu, S.~Canon, E.~A. Eloe-Fadrosh, {Anubhav}, M.~Babinski, Y.~Corilo,
  K.~Davenport, W.~D. Duncan, K.~Fagnan, M.~Flynn, B.~Foster, D.~Hays,
  M.~Huntemann, E.~K.~P. Jackson, J.~Kelliher, P.-E. Li, C.-C. Lo, D.~Mans,
  L.~A. McCue, N.~Mouncey, C.~J. Mungall, P.~D. Piehowski, S.~O. Purvine,
  M.~Smith, N.~J. Varghese, D.~Winston, Y.~Xu, and P.~S.~G. Chain, ``Challenges
  in {Bioinformatics} {Workflows} for {Processing} {Microbiome} {Omics} {Data}
  at {Scale},'' \emph{Frontiers in Bioinformatics}, vol.~1, Jan. 2022.
  [Online]. Available:
  \url{https://www.frontiersin.org/articles/10.3389/fbinf.2021.826370/full}
\BIBentrySTDinterwordspacing

\bibitem{clark_path_2019}
\BIBentryALTinterwordspacing
R.~D. Clark, ``A path to next-generation reproducibility in cheminformatics,''
  \emph{Journal of Cheminformatics}, vol.~11, no.~1, p.~62, Dec. 2019.
  [Online]. Available:
  \url{https://jcheminf.biomedcentral.com/articles/10.1186/s13321-019-0385-0}
\BIBentrySTDinterwordspacing

\bibitem{samuel_end--end_2022}
\BIBentryALTinterwordspacing
S.~Samuel and B.~König-Ries, ``End-to-{End} provenance representation for the
  understandability and reproducibility of scientific experiments using a
  semantic approach,'' \emph{Journal of Biomedical Semantics}, vol.~13, no.~1,
  p.~1, Dec. 2022. [Online]. Available:
  \url{https://jbiomedsem.biomedcentral.com/articles/10.1186/s13326-021-00253-1}
\BIBentrySTDinterwordspacing

\bibitem{goble_fair_2020}
\BIBentryALTinterwordspacing
C.~Goble, S.~Cohen-Boulakia, S.~Soiland-Reyes, D.~Garijo, Y.~Gil, M.~R. Crusoe,
  K.~Peters, and D.~Schober, ``{FAIR} {Computational} {Workflows},'' \emph{Data
  Intelligence}, vol.~2, no. 1-2, pp. 108--121, Jan. 2020. [Online]. Available:
  \url{https://www.mitpressjournals.org/doi/abs/10.1162/dint_a_00033}
\BIBentrySTDinterwordspacing

\bibitem{perezriverol_scalable_2020}
\BIBentryALTinterwordspacing
Y.~Perez-Riverol and P.~Moreno, ``{Scalable {Data}
  {Analysis} in {Proteomics} and {Metabolomics} {Using} {BioContainers} and
  {Workflows} {Engines}},'' PROTEOMICS,
  vol.~20, no.~9, p. 1900147, May 2020. [Online]. Available:
  \url{https://onlinelibrary.wiley.com/doi/10.1002/pmic.201900147}
\BIBentrySTDinterwordspacing

\bibitem{gruening_recommendations_2019}
\BIBentryALTinterwordspacing
B.~Gruening, O.~Sallou, P.~Moreno, F.~da~Veiga~Leprevost, H.~Ménager,
  D.~Søndergaard, H.~Röst, T.~Sachsenberg, B.~O'Connor, F.~Madeira,
  V.~Dominguez Del~Angel, M.~R. Crusoe, S.~Varma, D.~Blankenberg, R.~C.
  Jimenez, {BioContainers Community}, and Y.~Perez-Riverol,
  ``\BIBforeignlanguage{en}{Recommendations for the packaging and
  containerizing of bioinformatics software},''
  \emph{\BIBforeignlanguage{en}{F1000Research}}, vol.~7, p. 742, Mar. 2019.
  [Online]. Available: \url{https://f1000research.com/articles/7-742/v2}
\BIBentrySTDinterwordspacing

\bibitem{crusoe_methods_2022}
\BIBentryALTinterwordspacing
M.~R. Crusoe, S.~Abeln, A.~Iosup, P.~Amstutz, J.~Chilton, N.~Tijanić,
  H.~Ménager, S.~Soiland-Reyes, B.~Gavrilovic, and C.~Goble, ``Methods
  {Included}: {Standardizing} {Computational} {Reuse} and {Portability} with
  the {Common} {Workflow} {Language},'' \emph{Communications of the ACM},
  vol.~65, no.~6, pp. 54--63, Jun. 2022. [Online]. Available:
  \url{http://dx.doi.org/10.1145/3486897}
\BIBentrySTDinterwordspacing

\bibitem{ison_edam_2013}
\BIBentryALTinterwordspacing
J.~Ison, M.~Kalas, I.~Jonassen, D.~Bolser, M.~Uludag, H.~McWilliam, J.~Malone,
  R.~Lopez, S.~Pettifer, and P.~Rice, ``\BIBforeignlanguage{en}{{EDAM}: an
  ontology of bioinformatics operations, types of data and identifiers, topics
  and formats},'' \emph{\BIBforeignlanguage{en}{Bioinformatics}}, vol.~29,
  no.~10, pp. 1325--1332, May 2013. [Online]. Available:
  \url{https://academic.oup.com/bioinformatics/article-lookup/doi/10.1093/bioinformatics/btt113}
\BIBentrySTDinterwordspacing

\bibitem{gray_bioschemas_2017}
A.~Gray, C.~Goble, and R.~Jimenez, ``Bioschemas: {From} {Potato} {Salad} to
  {Protein} {Annotation},'' in \emph{Proceedings of the {ISWC} 2017 {Posters}
  \& {Demonstrations} and {Industry} {Tracks} co-located with 16th
  {International} {Semantic} {Web} {Conference} ({ISWC} 2017)}, 2017.

\bibitem{mondelli_bioworkbench_2018}
\BIBentryALTinterwordspacing
M.~L. Mondelli, T.~Magalhães, G.~Loss, M.~Wilde, I.~Foster, M.~Mattoso,
  D.~Katz, H.~Barbosa, A.~T.~R. de~Vasconcelos, K.~Ocaña, and L.~M. Gadelha,
  ``{BioWorkbench}: a high-performance framework for managing and analyzing
  bioinformatics experiments,'' \emph{PeerJ}, vol.~6, p. e5551, Aug. 2018.
  [Online]. Available: \url{https://peerj.com/articles/5551}
\BIBentrySTDinterwordspacing

\bibitem{babuji_parsl_2019}
\BIBentryALTinterwordspacing
Y.~Babuji, A.~Woodard, Z.~Li, D.~S. Katz, B.~Clifford, R.~Kumar, L.~Lacinski,
  R.~Chard, J.~M. Wozniak, I.~Foster, M.~Wilde, and K.~Chard, ``Parsl:
  {Pervasive} {Parallel} {Programming} in {Python},'' in \emph{28th {ACM}
  {International} {Symposium} on {High}-{Performance} {Parallel} and
  {Distributed} {Computing} ({HPDC})}, May 2019, \_eprint: 1905.02158.
  [Online]. Available: \url{http://dx.doi.org/10.1145/3307681.3325400}
\BIBentrySTDinterwordspacing

\bibitem{mondelli_exploring_2019}
\BIBentryALTinterwordspacing
M.~L. Mondelli, A.~Townsend~Peterson, and L.~M.~R. Gadelha, ``Exploring
  {Reproducibility} and {FAIR} {Principles} in {Data} {Science} {Using}
  {Ecological} {Niche} {Modeling} as a {Case} {Study},'' in \emph{Advances in
  {Conceptual} {Modeling}. {ER} 2019. {Lecture} {Notes} in {Computer}
  {Science}, vol. 11787}.\hskip 1em plus 0.5em minus 0.4em\relax Springer,
  2019, pp. 23--33. [Online]. Available:
  \url{http://link.springer.com/10.1007/978-3-030-34146-6_3}
\BIBentrySTDinterwordspacing

\bibitem{joachimiak_kg-microbe_2021}
M.~P. Joachimiak, H.~Hegde, W.~D. Duncan, J.~T. Reese, L.~Cappelletti, and
  A.~E. Thessen, ``{KG}-{Microbe}: a reference knowledge-graph and platform for
  harmonized microbial information,'' in \emph{The 12th {International}
  {Conference} on {Biomedical} {Ontologies} ({ICBO} 2021)}, 2021.

\end{thebibliography}
\end{document}